\documentclass[11pt, a4paper]{article}

\usepackage[utf8]{inputenc}
\usepackage[T1]{fontenc}
\usepackage[margin=1in]{geometry}
\usepackage[numbers]{natbib}
\usepackage{amsmath,amssymb,amsfonts,bm}
\usepackage{graphicx}
\usepackage{booktabs}
\usepackage{xcolor}
\usepackage{hyperref}
\usepackage{authblk} 

\newcommand{\copyrightnotice}{%
 \thanks{This is a Gold Open Access article made available under the CC-BY license. Published in \textit{IEEE Access}. DOI: \href{https://doi.org/10.1109/ACCESS.2025.3551798}{10.1109/ACCESS.2025.3551798}}
}

\title{\textbf{Energy-Efficient Prediction in Textile Manufacturing: Enhancing Accuracy and Data Efficiency With Ensemble Deep Transfer Learning}\copyrightnotice}

\author[1]{Yan-Chen Chen}
\author[2]{Wei-Yu Chiu\thanks{Corresponding author: chiuweiyu@gmail.com. }}
\author[1]{Qun-Yu Wang}
\author[3]{Jing-Wei Chen}
\author[3]{Hao-Ting Zhao}

\affil[1]{Department of Electrical Engineering, National Tsing Hua University, Hsinchu, Taiwan}
\affil[2]{School of Information Technology, Faculty of Science, Engineering and Built Environment, Deakin University, Victoria, Australia}
\affil[3]{Industrial Technology Research Institute, Hsinchu, Taiwan}

\date{} 

\begin{document}

\maketitle

\begin{abstract}
Traditional textile factories consume substantial energy, making energy-efficient production optimization crucial for sustainability and cost reduction. Meanwhile, deep neural networks (DNNs), which are effective for factory output prediction and operational optimization, require extensive historical data—posing challenges due to high sensor deployment and data collection costs. To address this, we propose Ensemble Deep Transfer Learning (EDTL), a novel framework that enhances prediction accuracy and data efficiency by integrating transfer learning with an ensemble strategy and a feature alignment layer. EDTL pretrains DNN models on data-rich production lines (source domain) and adapts them to data-limited lines (target domain), reducing dependency on large datasets. Experiments on real-world textile factory datasets show that EDTL improves prediction accuracy by 5.66\% and enhances model robustness by 3.96\% compared to conventional DNNs, particularly in data-limited scenarios (20\%–40\% data availability). This research contributes to energy-efficient textile manufacturing by enabling accurate predictions with fewer data requirements, providing a scalable and cost-effective solution for smart production systems.
\end{abstract}


\newpage
\section{Introduction}
\label{sec:introduction}
The textile industry is one of the most energy-intensive conventional industries, making energy-efficient production control essential for sustainability and cost reduction \cite{10Hong, 21Qaisar, 12hasanbeigi, 15hasanbeigi}. 
Among the various processes in textile manufacturing, the dyeing and finishing stage accounts for a significant proportion of energy consumption. One of the most energy-intensive machines in this process is the stenter setting machine (SSM), responsible for heat-setting fabric properties. However, the control of the heat-setting process has traditionally relied on workers' experience, leading to variations in quality and excessive energy consumption. 
Optimizing the control parameters of the heat-setting process can significantly reduce energy usage without compromising textile quality \cite{10Hong}. This study aims to develop an effective predictive model for optimizing the heat-setting process by integrating deep transfer learning (DTL) with ensemble learning, reducing data dependency while maintaining prediction accuracy.

In recent years, deep learning methods have been increasingly employed in industrial applications to improve energy efficiency \cite{23Fordal}. Deep neural networks (DNNs) have been successfully applied to model and predict industrial heat-drying processes \cite{15aghbashlo}, enabling better parameter optimization. 
Machine learning regression models and DNNs have also been used to analyze SSM \cite{19taur}, helping establish input--output relationships for drying processes in textile factories. However, most existing predictive models focus only on moisture content, ignoring other fabric quality indicators and energy consumption. Furthermore, these models often require complete and extensive historical data, which is costly and sometimes unavailable due to sensor limitations or production variations.

A major challenge in applying DNN models across production lines is the assumption that training and test data originate from the same distribution. In real-world scenarios, differences in machine characteristics, production modes, and environmental conditions introduce domain discrepancies, limiting model generalization \cite{13zhang,10Pan}. While one straightforward solution is to retrain models from scratch using newly collected data, this approach is time-consuming and costly, especially when historical data is sparse. Consequently, effective modeling for the heat-setting process must address cross-domain adaptation challenges and data limitations.

Deep transfer learning (DTL) provides a promising solution by enabling models trained in one domain to be adapted to another with minimal data \cite{18tan}. DTL has been successfully applied in energy-related fields. For example, Hu et al. \cite{16hu} transferred wind speed prediction models from data-rich wind farms to newly built farms. Ribeiro et al. \cite{18ribeiro} and Fan et al. \cite{20CFan} developed DTL-based approaches for cross-building energy forecasting. Since energy prediction tasks often involve time-series data, Long Short-Term Memory (LSTM) networks have been widely used for sequential modeling\cite{23Iftikhar}. Several studies have integrated LSTM with DTL for energy forecasting, such as thermal load prediction \cite{21Lu} and cross-building energy consumption \cite{21fang, 22Yusun}. Sarmas et al. \cite{22Sarmas} further improved solar power forecasting by incorporating transfer learning with stacked LSTM networks. 

Despite the success of DTL in energy-related applications, challenges remain in applying it to textile production. Excessive domain discrepancies or anomalous data can cause overfitting, leading to negative transfer \cite{22zhang}. Additionally, training LSTM-based DTL models requires complete time-series data, which is often unavailable in textile factories due to sensor malfunctions, communication errors, or environmental stress. Addressing these challenges requires a more robust transfer learning framework that can handle missing data and mitigate negative transfer effects.

Based on the literature review, further research is needed to improve predictive modeling for heat drying in textile manufacturing. First, existing models primarily focus on predicting moisture content, but comprehensive modeling should incorporate various fabric quality indicators and electricity consumption. Second, newly established production lines often lack sufficient historical data due to limited sensor equipment or recent sensor deployment. Although DTL is a viable solution, its effectiveness in complex factory environments must be evaluated, considering unstable data collection and domain adaptation issues.

To address these challenges, this study proposes an Ensemble Deep Transfer Learning (EDTL) approach that integrates deep transfer learning with ensemble learning techniques. Ensemble learning improves prediction accuracy by combining multiple models \cite{06polikar, 24iftikhar}. Prior studies \cite{20postel,20zhu,22Xia} have explored ensemble-based DTL to enhance stability in industrial settings. In our work, different layer-tuning strategies are applied to base learners, generating a diverse set of transferred models that are subsequently integrated using ensemble learning for final predictions.

The key contributions of this paper are as follows:
\begin{itemize}
    \item We propose an EDTL approach that addresses cross-production line domain differences and data scarcity. To the best of our knowledge, this is the first study to apply a DTL framework to textile manufacturing, demonstrating its feasibility with real-world production data. Our approach significantly reduces the required training data while maintaining high predictive accuracy.
    \item Unlike previous prediction models \cite{21Lu,21fang,22Yusun,22Sarmas}, our method does not rely on complete time-series data. Even with missing data, the ensemble learning component mitigates negative transfer, ensuring robust predictive performance under diverse conditions.
    \item Our framework supports multi-target prediction, allowing simultaneous estimation of electricity consumption and fabric quality indicators. This enables comprehensive process optimization to achieve both energy efficiency and high product quality in textile manufacturing.
\end{itemize}

To guide this research, we investigate the following key questions:
\begin{enumerate}
    \item Can deep transfer learning be effectively applied to cross-production line adaptation in textile factories?
    \item How does ensemble learning enhance the robustness and stability of DTL models under real-world data constraints?
    \item To what extent can EDTL reduce data dependency while maintaining predictive accuracy for textile heat-drying processes?
\end{enumerate}

The remainder of this paper is structured as follows.
 Section~\ref{rel_work} introduces related work. Section~\ref{sec_model} outlines the SSM system within the textile factory and the problem of interest. Section~\ref{sec_method} details the proposed method. Section~\ref{sec_experiment} presents the experimental results. Finally, conclusions are drawn in Section~\ref{sec_conclusion}.

\section{Related Work}\label{rel_work}
This section provides an overview of modeling techniques for the textile drying process and deep transfer learning approaches.

\subsection{Modeling Techniques for Textile Drying Process}
The drying process is a critical stage in the textile industry, significantly affecting fabric quality. To optimize fabric quality and energy efficiency, modeling techniques are essential for simulating and understanding the drying process \cite{22wilmer}. These techniques can be broadly categorized into physics-based models and data-driven models.

Physics-based models rely on thermodynamic principles and mathematical equations to describe heat and mass transfer. For instance, Santos et al. \cite{15santons} developed a heat and mass transfer model to estimate the environmental impact on SSM performance. Qian et al. \cite{19qian} analyzed the drying characteristics of cotton fabrics using a two-dimensional heat and moisture transfer model. Patel et al. \cite{21patel} investigated energy flows in the SSM manufacturing process and later refined their approach to derive energy efficiency parameters through mathematical formulations \cite{22patel}. While physics-based models provide valuable insights, they often fail to capture the implicit interactions between different process parameters, limiting their effectiveness in analyzing and controlling dynamic systems.

The emergence of big data and artificial intelligence has led to a growing preference for data-driven modeling in the textile industry \cite{18Yildirim}. Machine learning techniques offer a powerful means of handling highly complex and nonlinear processes and have achieved remarkable success in electricity price forecasting\cite{iftikhar2024electricity, shah2020modeling}. Akyol et al. \cite{150akyol} used feature selection techniques to determine the most influential manufacturing parameters, comparing various machine learning models for wool yarn bobbin drying. Their subsequent study \cite{151akyol} employed deep neural networks (DNNs) to predict drying rates and optimize drying conditions through thermodynamic analysis. Taur et al. \cite{19taur} introduced a stacked regressor for SSM analysis, demonstrating that DNNs provided superior predictive accuracy. However, data-driven models often struggle with distribution discrepancies and data scarcity\cite{an2024few}, which hinder their ability to generalize across different drying environments. Addressing these limitations requires advanced methods such as deep transfer learning (DTL).

\subsection{Deep Transfer Learning}
Transfer learning is a paradigm that enables models trained on one domain to be adapted for similar tasks with limited data \cite{10Pan}. Deep neural networks have achieved remarkable success in various fields, including medical diagnostics \cite{20Tang}, computer vision \cite{22Yu, 22Yanghui}, and industrial automation \cite{23Fordal}. Due to the success of deep learning in data-driven methods, deep transfer learning has gained popularity as a means of improving model performance while mitigating data scarcity issues. Existing DTL techniques can be categorized into Instance-based, Network-based, Adversarial-based, and Mapping-based methods \cite{18tan}.

Instance-based DTL focuses on reducing distribution differences between source and target domains by reweighting source instances \cite{19Wang, 21Mathelin}. This method indirectly increases data richness by supplementing target domain training with reweighted source data \cite{23He, 21Amirian}. However, its effectiveness is contingent upon a small distributional gap between domains, making it unsuitable when global conditional shifts occur in real-world settings.

Network-based DTL fine-tunes pre-trained networks on source data using target data \cite{14Yosinski,16hu,18ribeiro,20CFan,21Lu,21fang,22Yusun,22Sarmas}. The assumption underlying this approach is that models trained on related tasks can be efficiently adapted, thereby reducing training time and improving performance \cite{14Yosinski}. However, fine-tuning with limited target data risks overfitting, reducing generalization performance \cite{19Li}.

Adversarial-based DTL is inspired by Generative Adversarial Networks (GANs) \cite{16Ganin} and aims to learn domain-invariant feature representations. By ensuring that feature distributions remain indistinguishable between source and target domains, adversarial learning is highly effective for domain adaptation \cite{21fang,21Elnaz}. However, adversarial training is unstable when source and target datasets are imbalanced, leading to convergence issues \cite{22Zhao}.

Mapping-based DTL transforms source and target features into a shared latent space, reducing distributional discrepancies. Distance metrics such as Wasserstein distance \cite{18shen}, KL-Divergence \cite{19sun}, and Maximum Mean Discrepancy (MMD) \cite{14Tzeng,22wang} are commonly used to align feature distributions. Although MMD is widely employed, its computational complexity increases significantly with large datasets, and mapping features into a common space may lead to loss of crucial information.

\subsection{Summary and Research Rationale}
From the literature review, it is evident that physics-based models alone cannot effectively capture the dynamic and complex nature of the textile drying process. While data-driven models offer a powerful alternative, they are constrained by data scarcity and domain shifts, which limit their adaptability across different production environments. Deep transfer learning has emerged as a promising solution, yet existing approaches—particularly network-based DTL—face challenges such as overfitting due to limited target domain data. 

To address these challenges, this study proposes an Ensemble Deep Transfer Learning (EDTL) framework that integrates multiple network-based transfer learning models. The rationale for this approach is twofold: 1) leveraging ensemble learning enhances generalization and robustness, reducing overfitting risks, and 2) combining multiple pre-trained models enables knowledge transfer across different drying conditions, improving predictive accuracy. By applying EDTL to the textile heat drying process, we aim to develop a more generalizable and data-efficient predictive control system, optimizing energy usage while maintaining fabric quality.

\section{System Model and Problem of Interest}\label{sec_model}

\begin{figure*}[htb!]
\includegraphics[width=0.8\textwidth,clip,keepaspectratio]{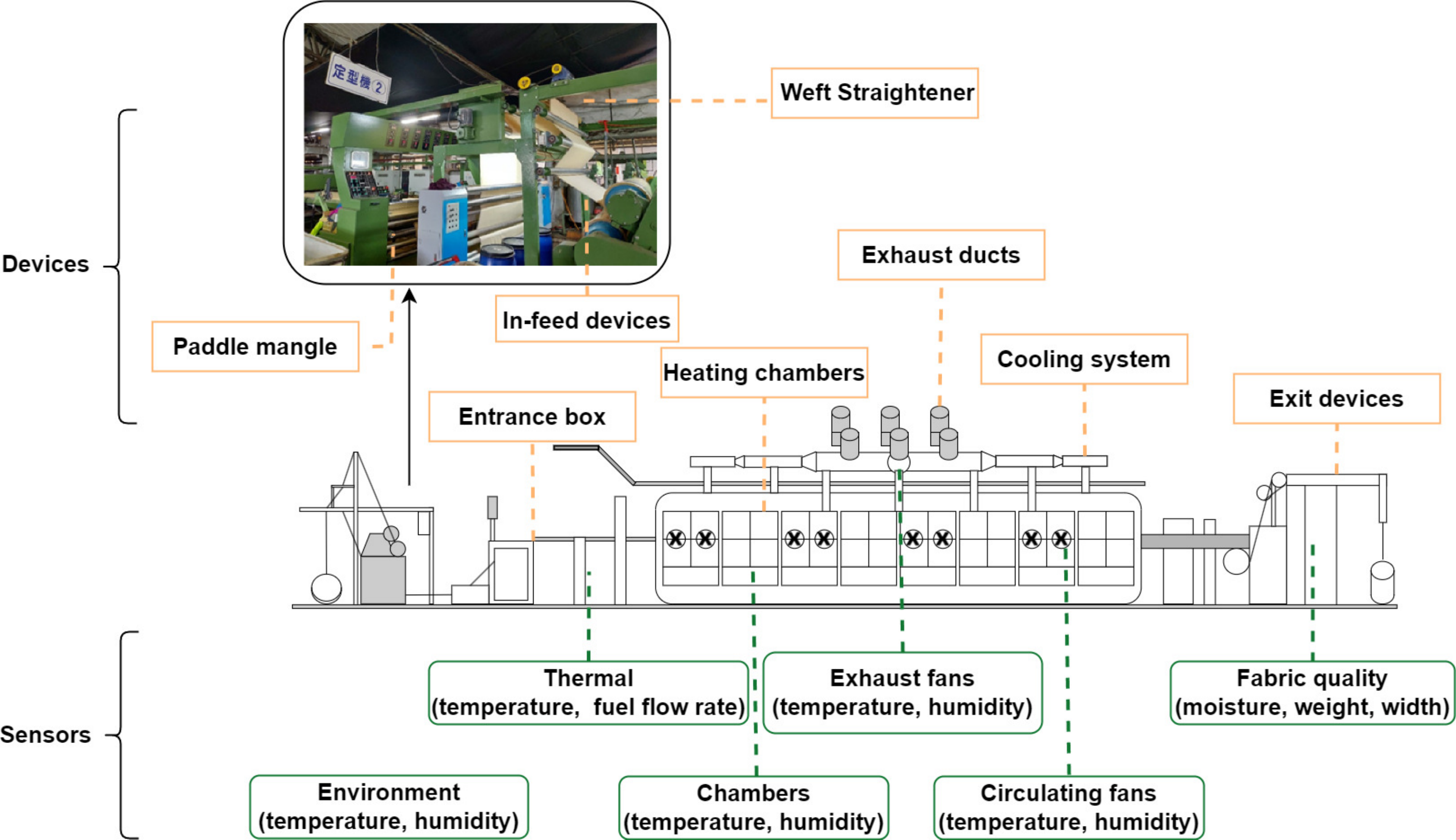}
\centering
\caption{Stenter Setting Machine (SSM).}\label{fig_SSM}
\end{figure*}

This section provides an overview of the SSM used in a textile factory for drying. The SSM is equipped with sensors that capture temporal data during its operations.
During the operation of the SSM, control parameters are adjusted and used to save energy consumption of the SSM. 
The prediction models forecast the impact of various control parameter settings. 
Nevertheless, domain differences between production lines restrict their scalability.
The problem of interest is then described.

\subsection{SSM in textile industry}
Producing finished fabrics in traditional textile manufacturing involves several stages, including pre-treatment, dyeing, and post-treatment. 
The post-treatment process consumes considerable energy.
In the post-treatment stage, the SSM shown in Figure~\ref{fig_SSM} is critical to stabilizing cloth and maintaining fabric quality without shrinkage.
The SSM is also responsible for applying finishes, drying, and heat-setting the fabrics through chamber heating, stretching, and sizing techniques\cite{22alam}.

Heat drying involves multiple heating chambers that use heat circulation and evaporation to remove moisture from the fabric. 
This process requires a significant amount of electricity, fuel, and steam to dry and shape the fabric\cite{21patel, 22dai}. In heat drying, the dominant contributors to electricity consumption, denoted by $E$, are the evaporation of water from the fabric and the heating of the air.

The quality of fabric produced on the production line is another crucial factor that must be considered. To evaluate the fabric quality, textile manufacturers often rely on three indicators: fabric weight ($W$), fabric width ($D$), and moisture content ($M$). The moisture content $M$ satisfies\cite{15baxi}
\begin{equation}
\begin{split}
\frac{dM}{dt}=\frac{1}{m_{\rm f}}[G_{\rm in}M_{\rm in}-G_{\rm out}M_{\rm out}-R_{\rm d}m_{\rm f}-M\frac{dm_{\rm f}}{dt}]
\end{split}
\end{equation}
where $m_{\rm f}$ is the fabric mass, $G_{\rm in}$ and $G_{\rm out}$ are the fabric mass flow input and output rates, respectively, and $R_{\rm d}$ is the drying rate. 
The drying rate $R_{\rm d}$ can be determined by
\begin{equation}
\begin{split}
R_{\rm d}=K(M-M_{\rm e})
\end{split}
\end{equation}
where $M_{\rm e}$ is the equilibrium moisture content of the fabric  (the moisture content that the fabric will reach when exposed to the same environmental conditions for a sufficient period of time), and $K$ is the drying constant expressed by\cite{15baxi}
\begin{equation}
\begin{split}
K=0.00719\exp({\frac{-130.64}{T_{\rm a}}})
\end{split}
\end{equation}
where $T_{\rm a}$ is the air temperature.

\subsection{Generating Optimal Control Parameters and Challenges}

The SSM is equipped with various sensors that monitor fabric quality, environmental changes, and control parameters. 
Fabric-related information includes fabric color, type, width, weight, and manufacturing process specifications, which is utilized to evaluate the quality of fabric products.
Environmental monitoring information is relevant to the changes in the operating environment of the SSM during production. Control parameters include  main motor speed, fan speed, and temperature setting. 

\begin{figure}[t]
\includegraphics[width=\columnwidth,clip,keepaspectratio]{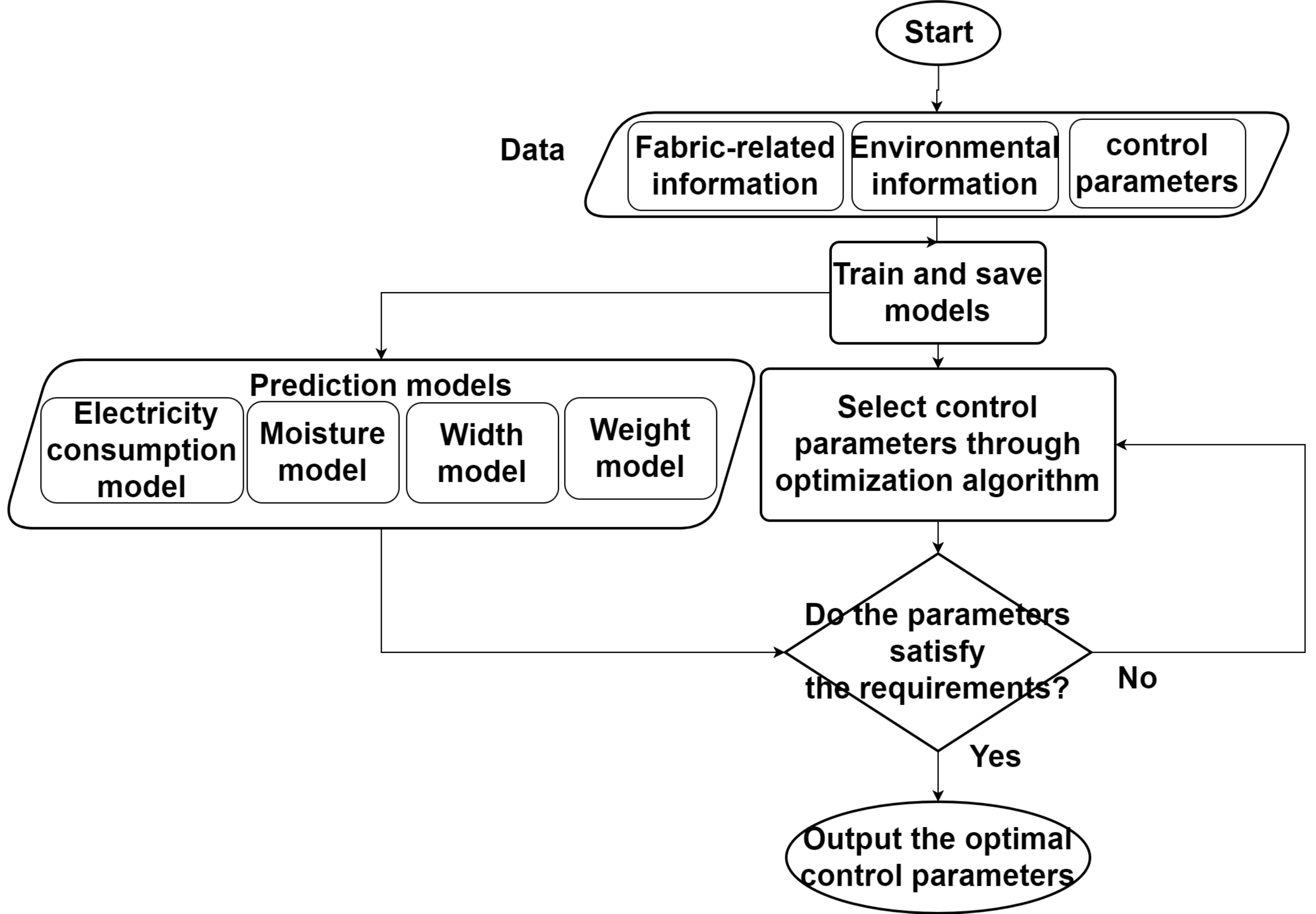}
\centering
\caption{Flowchart of generating control parameters.}\label{fig_process}
\end{figure}

With the data collected, the SSM system can be analyzed to identify the control parameters related to fabric quality and energy-saving goals.
Figure~\ref{fig_process} presents the flowchart of generating optimal control parameters.
First, temporal data is collected in the factory. 
The collected data is used to build prediction models that estimate fabric-related indicators' values, including electricity consumption $E$, moisture content $M$, fabric weight $W$, and fabric width $D$.

Based on these models, 
optimization algorithms can be applied to 
search for optimal control parameters for the production line in a SSM environment of the textile factory.
These optimized parameters are then used to ensure fabric quality and reduce 
energy consumption.

To determine optimal control parameters, both optimization algorithms and accurate prediction models are essential. While well-established optimization algorithms have been widely used, creating precise models in a SSM production line poses several challenges. For instance, the drying process of the setting machine is inherently complex, with physical and chemical properties that are constantly changing.

\subsection{Domain difference in cross-production line}

Sensor data is collected from various production lines through the factory monitoring system platform. Based on the richness of historical information, it can be divided into the source and target domains. 

Let $\mathcal{S}=\{(x_i^{\rm s}, y_i^{\rm s})\}_{i=1}^{N^{\rm s}}$ represent the datasets from the source domain, where $N^{\rm s}$ denotes the number of samples, $x_i^{\rm s}$ represents a multi-dimensional input feature, and $y_i^{\rm s}$ is the corresponding output with continuous regression labels. More specifically, $x_i^{\rm s}$ can represent environmental information and control parameters related to SSM, and $y_i^{\rm s}$ can represent the predicted targets such as energy consumption $E^{\rm s}$, fabric weight $W^{\rm s}$, fabric width $D^{\rm s}$, and fabric moisture content $M^{\rm s}$. Similarly, the target domain dataset can be defined as $\mathcal{T}=\{(x_i^{\rm t},y_i^{\rm t})\}_{i=1}^{N^{\rm t}}$, where $N^{\rm t}$ is the number of samples in the target domain, comprising input features $x_i^{\rm t}$ and response outputs $y_i^{\rm t}$ (i.e., $E^{\rm t}$, $W^{\rm t}$, $D^{\rm t}$, and $M^{\rm t}$). 
Since the source domain has richer historical data, $N^{\rm s}$ is significantly larger than $N^{\rm t}$.

In general, the prediction model is trained using data from the source domain, and evaluated on target domain data. However, the effectiveness of this method depends on whether the data of the two domains follow the same distribution. Unfortunately, in real-world industrial scenarios, there are variations in machine characteristics or production modes across different production lines~\cite{20Azamfar}. In the textile industry, these variations can occur due to a) changes in the equipment (SSM and sensors from various manufacturers or models, resulting in variations in performance, precision, and operating speed), b) changes in the process or product specification (the update of manufacturing processes or fabric specifications can alter the way equipment operates, such as temperature control, speed settings, or processing sequences). These variations lead to a mismatch in sensor data distribution, further causing the domain shift:
\begin{equation}
        P(y_{\rm s}|x_{\rm s}) \neq P(y_{\rm t}|x_{\rm t})
    \label{domain shift}
\end{equation}
where $\{y_{\rm s},x_{\rm s}\} \in \mathcal{S}$, and $\{y_{\rm t},x_{\rm t}\} \in \mathcal{T}$. Therefore, using the model trained with source domain data to directly predict the target domain data may result in reduced generalization capability.

In addition, the feature representations in different domains may vary. Let $\mathcal{F}^{\rm s}$ and $\mathcal{F}^{\rm t}$ can be denoted the feature sets in the source and target domains, respectively.
Owing to different monitoring needs and costs, the number and types of sensors deployed across different production lines are various, leading to distinct input feature spaces: 
\begin{equation}
        \mathcal{F}^{\rm s} \neq \mathcal{F}^{\rm t}
        \label{distinct feature space}
\end{equation}
Since the input space of the model cannot be arbitrarily adjusted to varying data features, this restricts its application across different domains.

To solve these problems, rebuilding models from scratch using newly collected data is a straightforward solution. However, installing various sensor devices can be costly, especially in conventional textile factories that rely on more standardized sensors and monitoring systems.
In addition, the data collection process within a factory can be time-consuming. Factories that have recently installed sensors may only have short-term data collection; as such, data may not accurately represent the entire range of operating conditions, yielding overfitting or prediction bias in prediction models.


\section{The workflow for constructing model and proposed Ensemble deep transfer learning}\label{sec_method}

This section describes the process for developing various prediction models, including one energy consumption model (electricity) and three fabric quality models (moisture model, width model, and weight model).
By utilizing transfer learning, prediction models can leverage sensor information from similar production lines to enhance prediction accuracy. 
However, using transfer learning exclusively to develop prediction models still results in large prediction errors if there is significant variability in data distributions of different factories.   
An ensemble learning method is thus integrated with transfer learning models to minimize the prediction bias and further improve model accuracy.

\begin{figure}[t]
\includegraphics[width=\columnwidth,clip,keepaspectratio]{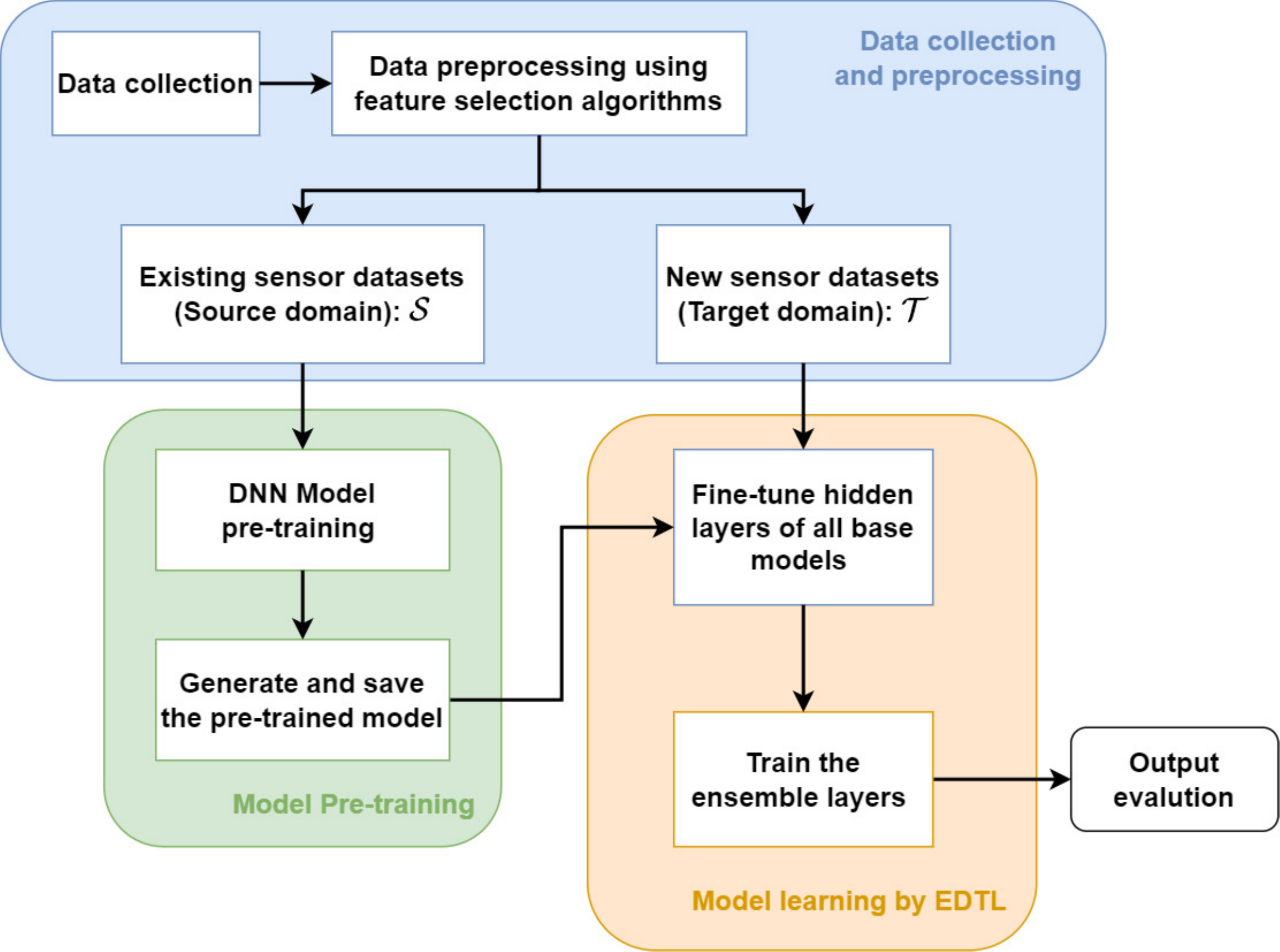}
\centering
\caption{Proposed workflow for constructing prediction models. }\label{fig_workflow}
\end{figure}

Figure~\ref{fig_workflow} 
presents the workflow
to build the prediction models in Figure~\ref{fig_process}. The workflow can be divided into three stages: 1) collection and preprocessing of temporal data in a textile factory, 2) pre-training of the source domain model, and 3) model learning by EDTL in the target domain, that is our proposed method.

\subsection{Data Collection and preprocessing}
The process data of SSM is collected by sensors equipped on production lines. Usually, the source domain data comes from the previous production lines, that have diverse historical data available for analysis and modeling. In contrast, the target domain data collected from newly established production lines, due to its poor data.

After data collection, it is pre-processed by eliminating redundant or missing ones. 
To identify the most influential features for a model, some feature selection  algorithms can be applied.  Selected features are then used to learn the prediction models.

\subsection{Model Pre-training}
The pre-trained model is created using a DNN, and its training is based on data from $\mathcal{S}$. 
Each layer within the network produces an output by applying weights and biases to the inputs and passing the resulting values through an activation function. 
These outputs then serve as inputs for the subsequent layer, establishing a coherent flow throughout the network.

In a DNN model, the output $h_{(j)}$ of each layer can be calculated as
\begin{equation}
h_{(j)}=g(\boldsymbol{W}_{(j)}h_{(j-1)}+b_{(j)}) \quad \quad {\forall} j {\in} \{1,2,...,J\}
\end{equation}
where $\boldsymbol{W}_{(j)}$ and $b_{(j)}$ represent the weight and bias in the $j$th hidden layer, respectively,  $g$ is the activation function, and $J$ represents the number of hidden layers.

The parameters of the DNN  $\theta=\{(\boldsymbol{W}_{(j)},b_{(j)})\}^{J}_{j=1}$ 
can be obtained   by minimizing 
loss function $L(\theta)$:  
\begin{equation}\label{loss_function}
\begin{split}
L(\theta)=\frac{1}{N}\sum_{i=1}^{N}(y_{i}-f(x_{i};\theta))^2=\frac{1}{N}\sum_{i=1}^{N}(y_{i}-\hat{y_{i}})^2
\end{split}
\end{equation}
where $x_{i}$ and $y_{i}$ represent the input and output of the training data, respectively, $N$ represents the number of training data, where $N <N^{\rm s}$, and $\hat{y_{i}}$ symbolizes the predicted value of the DNN model output using the weight and bias parameter $\theta$. The function $f$ encompasses the entire DNN model, including the specific architecture, activation functions, and weights.

During the training, the parameters of all network layers are updated to minimize the loss function. In (\ref{loss_function}), the assignments $\theta=\theta^{\rm s}$, $x_i=x_i^{\rm s}$, $y_i=y_i^{\rm s}$, and $N=N^{\rm s}$ ensure that the updates are performed using the corresponding values from the source domain.
Upon completion, the model $\theta^{\rm s}$ is saved in the pre-training phase and then transferred to the next stage.

\begin{figure}[htbp]
\includegraphics[width=\columnwidth,clip,keepaspectratio]{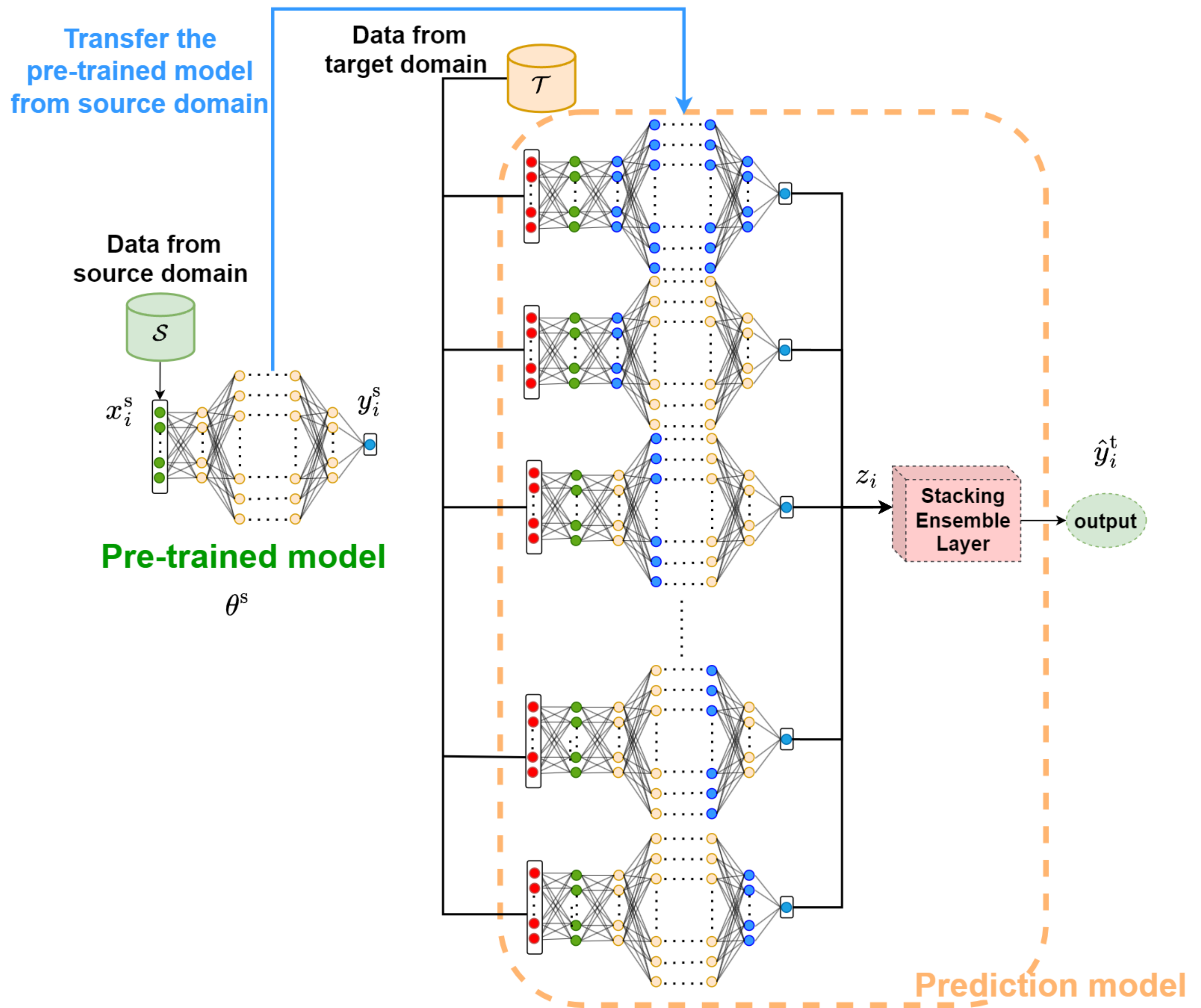}
\centering
\caption{Proposed EDTL.}\label{fig_EDTL}
\end{figure}

\subsection{Proposed Ensemble deep transfer learning}

The prediction models are constructed using the proposed EDTL approach, which combines the concept of transfer learning and ensemble learning. A prediction model is comprised of multiple base models, each of which inherits knowledge from the pre-trained model, and then are integrated by ensemble method. The details of EDTL will then be described.

After pre-training, the knowledge of source domain is transferred by network-based deep transfer learning\cite{18tan}. The technique of network-based deep transfer learning involves reusing the partial network pre-trained in the source domain, including its network structure and layer parameters. However, in the scenario of changed input feature (constraint \eqref{distinct feature space}), directly reusing the network is impractical, due to the unequal input dimensions, i.e. $dim(x_i^{\rm s}) \neq dim(x_i^{\rm t})$. To address this, we introduce a new layer with randomly initialized weights $w^0$ and biases $b^0$ while reusing the network. This additional layer helps adapt the model to the specific input dimensions of the target domain data.

Subsequently, the objective is to generate an ensemble framework. To develop an outstanding ensemble model, two conditions should be satisfied: 1) 
a diverse base model set; 2) an effective integration strategy. Through deep transfer learning, reusing and fine-tuning network layer parameters allows the model to adapt across different domains. Further,  fine-tuning different network layer can enable the model to learn diverse information within the same data. Given this opinion, in our method, multiple base models sharing the same network structure are created, each partially inherits parameters from the pre-trained model. To promote diversity among the base models, only one hidden layer in each network is fine-tuned, while the remaining layers are frozen. Additionally, a base model with all layers fine-tuned is included in the base model set, avoiding significant bias. 

Finally, an ensemble method integrates the outputs of all base models to obtain the optimal prediction result. Stacking regression\cite{06polikar}, an ensemble learning technique, is utilized to combine multiple base regressors via a meta-regressor. In this study, Support Vector Regression (SVR)\cite{15awad} is adopted as the meta-regressor, mapping the input to a high-dimensional space using a nonlinear mapping function and performing linear regression in that space. SVR effectively mitigates overfitting by incorporating regularization terms, while its convex optimization problem ensures a globally optimal solution.

During the training of EDTL, the parameters of all base models are updated to minimize the loss function. 
In (\ref{loss_function}),  $\theta=\theta^{\rm t}$, $x_i=x_i^{\rm t}$, $y_i=y_i^{\rm t}$, and $N=N^{\rm t}$, utilizing the corresponding values.
The output vector $z_i$ of all base models can be expressed as
\begin{equation}
\begin{split}
z_i = [ \ f_1(x_i^{\rm t};\theta_1^{\rm t}),\ f_2(x_i^{\rm t};\theta_2^{\rm t}),\ ...,\ f_{\rm k}(x_i^{\rm t};\theta_{\rm k}^{\rm t}) \ ]
\end{split}
\end{equation}
where $\rm k$ is the total number of base models. The estimated output $\hat{y}_i^{\rm t}$
can be calculated as 
\begin{equation}
\begin{split}
\hat{y}_i^{\rm t}=w^\top\phi(z_i)+b^{\rm svr} 
\end{split}
\end{equation}
where $\phi(z_i)$ is a nonlinear function used to map $z_i$ to a high-dimensional feature, and $w$ and $b^{\rm svr}$ are the weight of the feature space and the bias term in the regression, respectively.

In nonlinear function, a commonly used approach is to utilize a kernel function. The Gaussian radial basis function kernel in SVR is a popular choice among the various options discussed in the literature. This kernel can be expressed as
\begin{equation}
K(z,z_i)=\exp({-\frac{||z-z_i||^2}{2\gamma^2}}).
\end{equation}
Through the application of this kernel function, the corresponding predicted value $\hat{y}_i^{\rm t}$ can be represented as
\begin{equation}\label{kernel_output}
\hat{y}_i^{\rm t}=\sum_{i=1}^{N^{\rm t}}(\xi_i-\xi_i^*)K(z,z_i)+b^{\rm svr}
\end{equation}
where $z$ represents the variable of the input vector used in the kernel function, and the kernel function $K(z,z_i)$ calculates the distance between the variable $z$ and the training data $z_i$.

The objective of SVR is to discover a hyperplane that maximizes the margin between the predicted values $\hat{y}_i$ and the actual values $y_i$ while minimizing the prediction errors $\varepsilon$ that occur outside the margin boundaries $\xi_i$ and $\xi_i^*$.
To achieve this objective, the following optimization problem can be solved:
\begin{equation}\label{svr_equation}
\begin{split}
{\min}_{w,c,\xi_i, \xi_i^*}\frac{1}{2}w^{\top}w+C\sum_{i=1}^{N^{\rm t}}(\xi_i+\xi_i^*)
\end{split}
\end{equation}
subject to
\begin{equation*}
\begin{split}
\hat{y}_i-w^{\top}z_i-b^{\rm svr}\leq\varepsilon+\xi_i
\end{split}
\end{equation*}
\begin{equation*}
\begin{split}
w^{\top}z_i+b^{\rm svr}-\hat{y}_i\leq\varepsilon+\xi_i^*
\end{split}
\end{equation*}
\begin{equation*}
\begin{split}
\xi_i, \xi_i^*\geq0, i=1,2,...,N^{\rm t}
\end{split}
\end{equation*}
where $C$ is a hyperparameter that controls the tradeoff between the margin and error to avoid overfitting.

Figure~\ref{fig_EDTL} shows the framework of the EDTL. Firstly, a pre-trained model is transferred to the target domain. A few base models are produced by fine-tuned different hidden layers. In Figure~\ref{fig_EDTL}, the orange nodes in a hidden layer represent the frozen parts, the blue nodes are the parts that undergo fine-tuning in training, and the red nodes represent the newly introduced layer. Stacking Ensemble layer denotes using the ensemble method to integrate the outputs $z_i$ of all base models.

\begin{figure}[t]
\includegraphics[width=\columnwidth,clip,keepaspectratio]{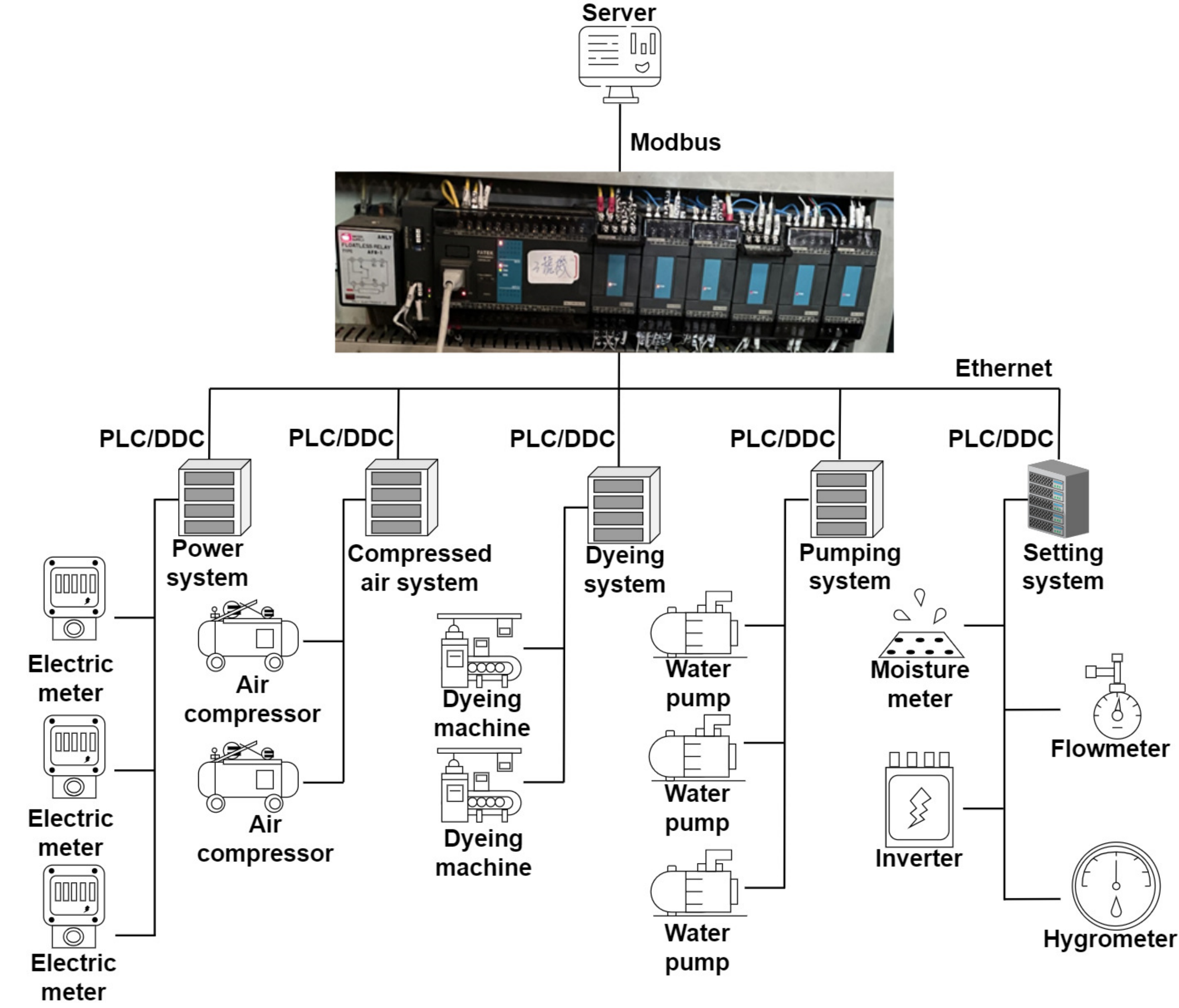}
\centering
\caption{Data collection platform in textile factories.}\label{fig_data}
\end{figure}

\section{Experimental Results}\label{sec_experiment}
This section first explains the data collection process, as well as the training process and configuration of the model. Next, experiments are designed based on different scenarios and compared with baselines to validate the effectiveness of the proposed method. The experiments primarily test the predictive performance of the trained model under different data availability cases and consider the impact of two scenarios on model learning, including the quality of the pre-trained model and the interference of anomalous data. Moreover, we conducted comparisons of electricity consumption with other existing models.

\subsection{Data collection and experimental design}

The experimental data was collected from real textile factories using an advanced monitoring system, as illustrated in Figure~\ref{fig_data}. This system enables real-time tracking of sensor measurements installed in the SSM equipment. Data transmission was carried out via wired communication, utilizing Ethernet and Modbus protocols (RS232, RS485, and TCP). The monitoring infrastructure consists of programmable logic controllers (PLCs) and Direct Digital Control (DDC) controllers, ensuring seamless data acquisition. SQL serves as an intermediary for efficient data exchange between platforms and monitoring systems.

\begin{figure}[!ht]
\includegraphics[width=\columnwidth,clip,keepaspectratio]{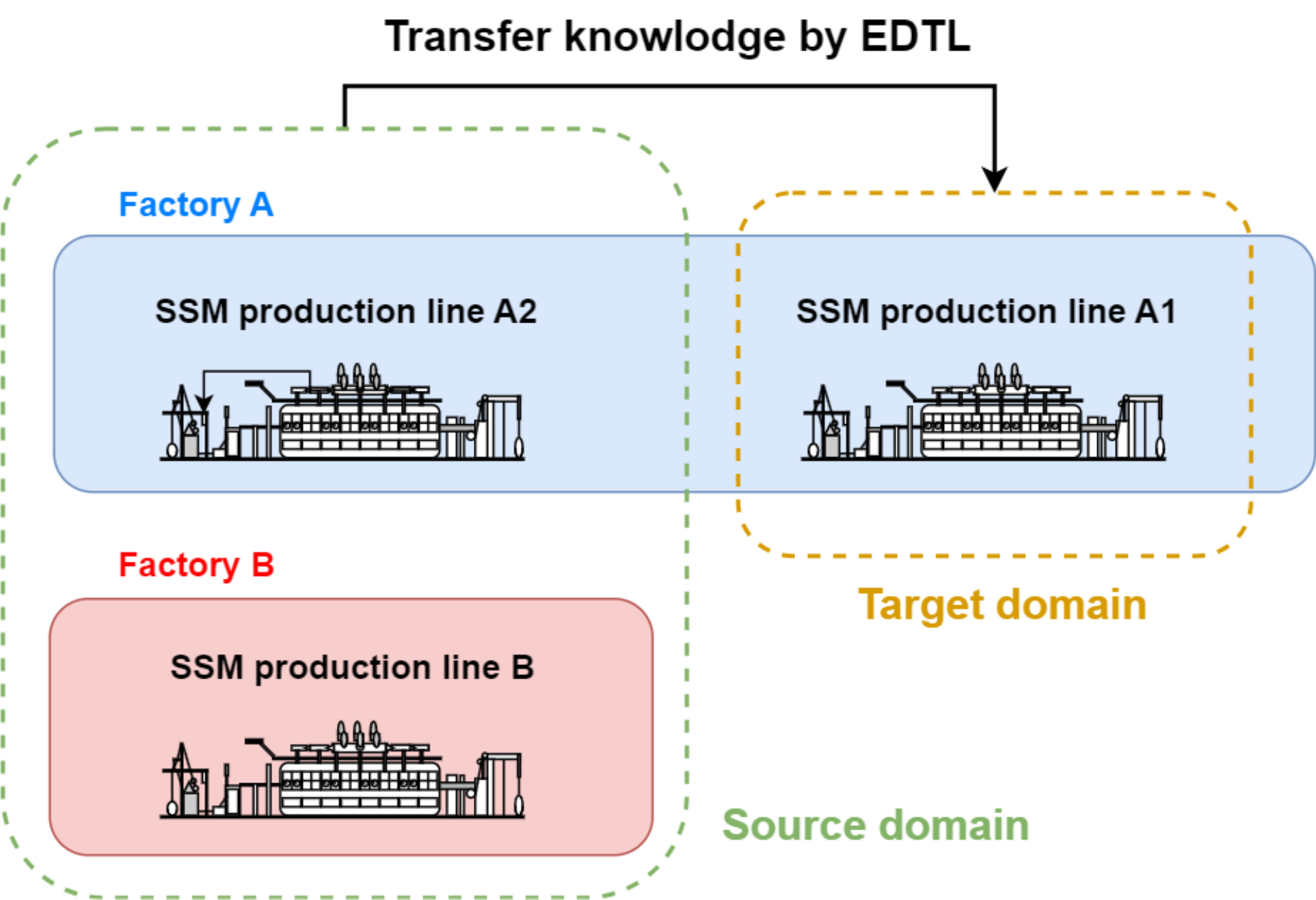}
\centering
\caption{Production lines in the source and target domains.}\label{production_lines}
\end{figure}

Real-time operational data was gathered from three SSM production lines: Production Line 1 of Factory A (denoted as A1), Production Line 2 of Factory A (denoted as A2), and the sole production line of Factory B (denoted as B). The datasets differ across production lines due to variations in sensor configurations. Key recorded features include fabric-related attributes, environmental monitoring data, and control parameters.  
Fabric-related attributes encompass fabric color, type, width, weight, moisture content, and process specifications.  
Environmental monitoring data include chamber temperature, humidity, real-time heat transfer fluid flow, multiple cloth temperature readings, and various thermal measurements.  
Control parameters cover main motor speed, chamber fan speeds, exhaust speeds, and temperature settings.  
Figure~\ref{production_lines} provides an overview of the production lines within the source and target domains. Given that A2 and B contain richer datasets with a greater number of features, they are designated as source domains. Conversely, A1, which has relatively limited data, serves as the target domain.  

To ensure data quality, preliminary preprocessing was conducted, including the removal of null values from the sensor data. The volume of cleaned data is summarized in Table~\ref{table_amount}.

\begin{table}[t]
\centering
\caption{Data Quantities}
\label{table_amount}
\begin{tabular}{cccc}
\toprule
Line&Nylon&Polyester&Total\\
\midrule
A1&6,139&6,550&12,689\\
\midrule
A2&10,386&14,356&24,742\\
\midrule
B&-&151,996&151,996\\
\bottomrule
\end{tabular}
\end{table}

Before training, we employed feature selection algorithm \cite{20kuzlu} to identify the most significant data features contributing to prediction target. To pre-train the DNN models, we divided the data in the source domain into a training set comprising 90\% of the data and a validation set containing the remaining 10\%. Pre-trained models had six layers, consisting of an input layer, an output layer, and four hidden layers.
Rectified linear unit (ReLU) activation functions were utilized for all the hidden layers except for the output layer; the output layer was designed without any activation function, allowing the network's output to be directly generated as the final result. The network was trained by the Stochastic Gradient Descent(SGD) algorithm with 30 epochs, and the batch size was set to 64. To achieve global optima, the Adam optimizer was adopted to perform dynamic adjustment, and the defaulted learning rate is 0.001.

\if
Using the Adam optimizer,parameter $\theta$ can be updated iteratively according to 
\begin{equation}
\begin{split}
\hat{m_t}=\frac{m_t}{1-\beta_1^t},\hat{v_t}=\frac{v_t}{1-\beta_2^t}
\end{split}
\end{equation}

\begin{equation}
\begin{split}
\theta_{t}=\theta_{t-1}-\frac{\gamma\hat{m_t}}{\sqrt{\hat{v_t}}+\epsilon}
\end{split}
\end{equation}

where $\hat{m_t}$ denotes the bias corrected estimator for the first moment $m_t$ at time $t$, $\hat{v_t}$ denotes the bias corrected estimator for the second moment $v_t$ at time $t$, $\gamma$ denotes the learning rate(or the step size), and $\beta_1,\beta_2$  are exponential decay rate for the moment estimates.
\fi

To learn the prediction models using the proposed EDTL in target domain, we utilized a month of data from production line A1 as the test dataset. The prediction models contained five base models, each of which had an architecture consisting of six layers. 
One layer involved a new input layer transforming inputs of varying sizes.
Based on the pre-trained models, we formed five base models by incrementally freezing different hidden layers, including one based model without any. 
After training each base model independently, the ensemble method was applied to determine the final prediction results. 
The remaining training settings in target domain were identical to those in the pre-training process.

\begin{table}[!ht]
\centering
\caption{Performance of Pre-trained Models in MAPE(\%)}
\label{table_pretrain}
\begin{tabular}{ccccc}
\toprule
Line-Type&Electricity $E^{\rm s}$&Moisture $M^{\rm s}$&Width $D^{\rm s}$&Weight $W^{\rm s}$\\
\midrule
A2-Nylon&5.10 &2.34 &1.38 &5.49 \\
\midrule
A2-Polyester&3.79 &2.88 &0.76 &8.14 \\
\midrule
B-Polyester&2.84 &0.40  &9.73 &7.19 \\
\bottomrule
\end{tabular}
\end{table}

\begin{figure}[!ht]
    \centering
    \begin{minipage}{0.80\linewidth}
        \centering
        \includegraphics[width=\linewidth]{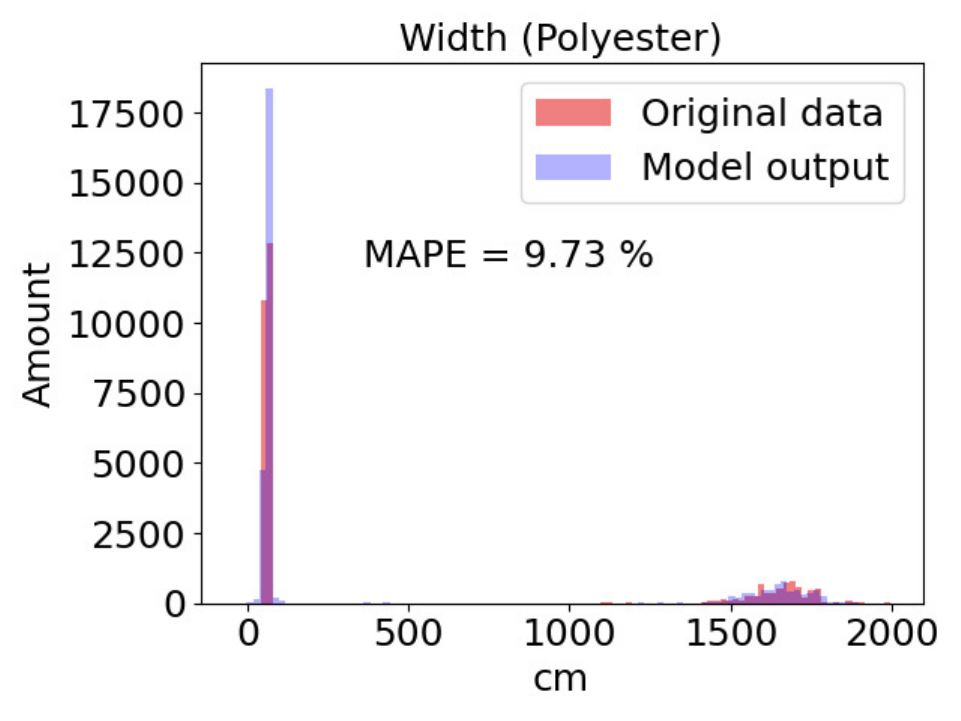}
        \centerline{(a)} 
        \label{subfig:a}
    \end{minipage}

    \vspace{0.5cm} 

    \begin{minipage}{0.80\linewidth}
        \centering
        \includegraphics[width=\linewidth]{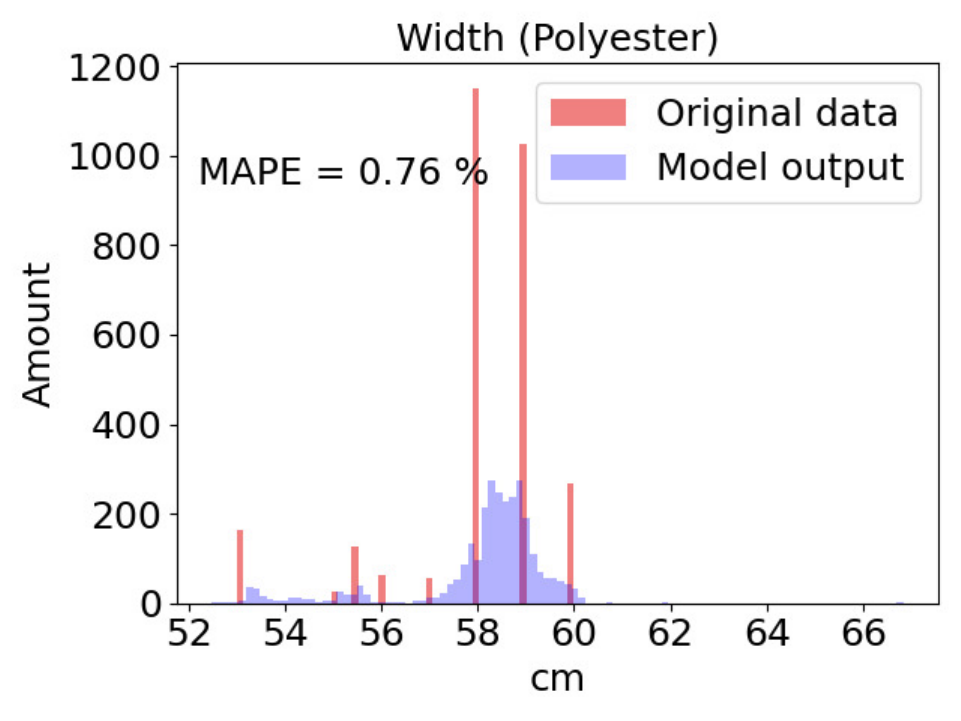}
        \centerline{(b)} 
        \label{subfig:b}
    \end{minipage}

   \caption{(a) Prediction of width (Polyester) distribution in B;(b) Prediction of width (Polyester) distribution in A2.}
 	\label{comparison_source}
\end{figure}

\begin{figure*}[!ht]
    \centering
        \includegraphics[width=\linewidth]{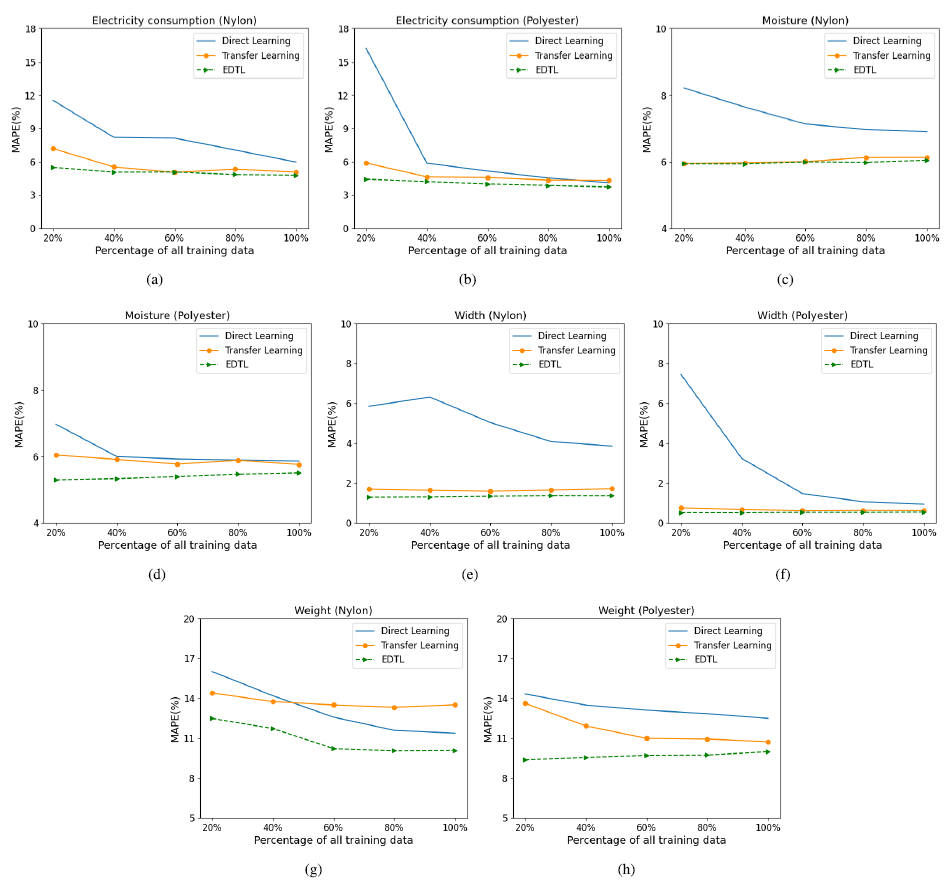}
    \caption{MAPE variation of different data usage (from A2 to A1). (a) Electricity consumption (Nylon); (b) Electricity consumption (Polyester); (c) Moisture (Nylon); (d) Moisture (Polyester); (e) Width (Nylon); (f) Width (Polyester); (g) Weight (Nylon); (h) Weight (Polyester).}
    \label{comparison}
\end{figure*}

To evaluate the prediction performance for $E^{\rm t}$, $W^{\rm t}$, $D^{\rm t}$, and $M^{\rm t}$, the mean absolute percentage error (MAPE) was considered:
\begin{equation}\label{eq_mape} \text{MAPE} = \frac{1}{N^{\rm t}}\sum_{i=1}^{N^{\rm t}}\left|\frac{{y_i}-\hat{y}_i}{y_i}\right|\times100\%.
\end{equation}
This metric calculates the average percentage difference between the predicted values $\hat{y}_i$ and the actual values $y_i$ for a total of $N^{\rm t}$ data points.

\subsection{Performance of pre-trained models in source domain}

Table~\ref{table_pretrain} presents the prediction performance for two factories.  The pre-trained models show strong prediction abilities, with acceptable prediction errors. Hence, these pre-trained models are suitable for transfer learning.
However, in the polyester case, the predictive performance for width in production line B was worse to that in production line A2 by a significant margin. This can be attributed to a substantial difference in the actual width distribution between the two production lines. These prediction outcomes are shown in Figure~\ref{comparison_source}(a) and (b).

\begin{figure}[!ht]
\includegraphics[width=\linewidth,clip,keepaspectratio]{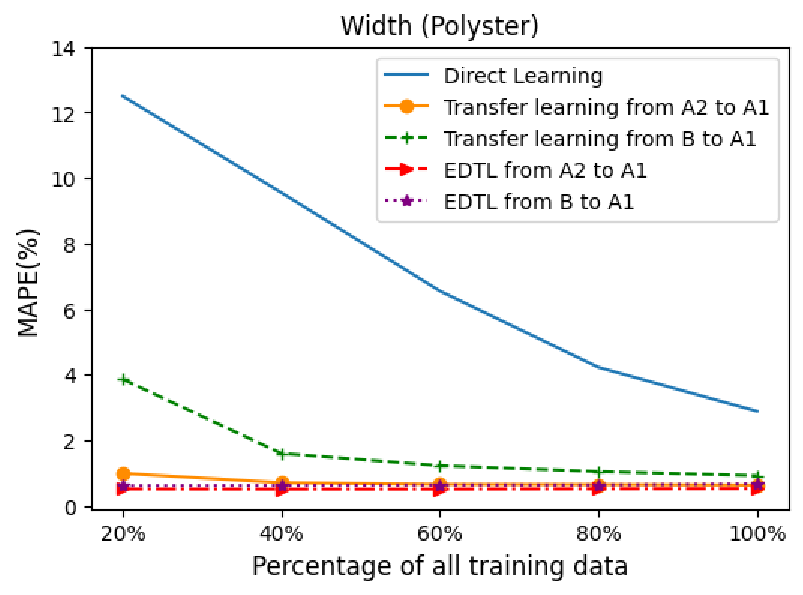}
\centering
\caption{Width prediction of transferring pre-trained models from two source domains.}\label{fig_comparison}
\end{figure}

\subsection{Analysis of prediction performance of EDTL
}

In the experiments, the pre-trained model from line A2 was utilized for knowledge transfer. Subsequently, the target domain model was fine-tuned using different proportions(20\%, 40\%, 60\%, 80\%, and 100\%) of line A1 data to predict four targets, including $E^t$, $M^t$, $W^t$, and $D^t$. The analysis was conducted separately for the production conditions of two types of fabrics, Nylon and Polyester. 

To demonstrate the advantages of the EDTL method, we included two baseline models for comparison: 1) Direct Learning (an initialized neural network trained from scratch on the target domain data directly), b) Transfer Learning (a single pre-trained neural network was fine-tuned using target domain data).

Figure~\ref{comparison} shows the prediction performance in terms of MAPE for eight different cases.
The results show that as data completeness increases, training models with neural networks can significantly reduce prediction errors. However, incorporating transfer learning resulted in models that exhibited lower prediction errors, even with limited data available. 
Based on all the results from Figure~\ref{comparison}, it can be observed that transfer learning achieved satisfactory predictive performance even with a small amount of data (20\%-40\%), while EDTL further enhanced the performance due to its integration with ensemble methods.

The following conclusions can be drawn. Direct learning in production line A1 (target domain) gradually achieved satisfactory prediction performance upon the increase in data usage; unfortunately, it may not be suitable for applying this method to a newly constructed production line having little data collected. By contrast, transfer learning using the pre-trained models can significantly reduce the sensitivity of neural networks to the amount of data required (40\% of data from the target domain were sufficient to attain accurate prediction models in the experiment). This method enables the new production lines without much data collection to determine operation parameters in consideration of electricity consumption and textile quality. Furthermore, EDTL integrates ensemble methods by stacking transferred base learners and making collective decisions, further improving performance.

\begin{table*}[!ht]
\centering
\footnotesize
\caption{Comparison of MAPE for electricity consumption prediction}
\label{Performance degradation (electricity)}
\setlength{\tabcolsep}{8pt} 
\begin{tabular}{llccccc}
\toprule
\textbf{Fabric} & \textbf{Model} & \multicolumn{5}{c}{\textbf{Percentage of Training Data}} \\
\cmidrule(lr){3-7}
\textbf{Type} & & \textbf{20\%} & \textbf{40\%} & \textbf{60\%} & \textbf{80\%} & \textbf{100\%} \\
\midrule
\textbf{Nylon} & Direct Learning & 27.46 / 11.53 & 31.99 / 8.20 & 33.88 / 7.05 & 32.60 / 7.05 & 32.11 / 5.97 \\
& & (+15.93) & (+23.79) & (+25.75) & (+25.55) & (+26.14) \\
\cmidrule{2-7}
& Transfer Learning & 31.04 / 7.18 & 30.06 / 5.53 & 27.09 / 5.03 & 24.01 / 5.32 & 21.59 / 5.06 \\
& & (+23.86) & (+24.53) & (+22.06) & (+18.69) & (+16.53) \\
\cmidrule{2-7}
& EDTL & \textbf{20.14 / 5.49} & \textbf{18.34 / 5.05} & \textbf{17.91 / 5.06} & \textbf{18.12 / 4.84} & \textbf{17.59 / 4.78} \\
& & \textbf{(+14.65)} & \textbf{(+13.29)} & \textbf{(+12.85)} & \textbf{(+13.28)} & \textbf{(+12.81)} \\
\midrule
\textbf{Polyester} & Direct Learning & 25.65 / 16.23 & 18.18 / 5.87 & 17.93 / 5.16 & 16.97 / 4.53 & 15.76 / 4.10 \\
& & (+9.42) & (+12.31) & (+12.77) & (+12.44) & (+11.66) \\
\cmidrule{2-7}
& Transfer Learning & 18.66 / 5.88 & 17.38 / 4.63 & 16.27 / 4.58 & 14.39 / 4.34 & 12.82 / 4.32 \\
& & (+12.78) & (+12.75) & (+11.69) & (+10.05) & (+8.50) \\
\cmidrule{2-7}
& EDTL & \textbf{13.82 / 4.43} & \textbf{12.49 / 4.20} & \textbf{11.98 / 3.99} & \textbf{11.27 / 3.86} & \textbf{10.62 / 3.72} \\
& & \textbf{(+9.39)} & \textbf{(+8.29)} & \textbf{(+7.99)} & \textbf{(+7.41)} & \textbf{(+6.90)} \\
\bottomrule
\addlinespace[1ex]
\multicolumn{7}{l}{\footnotesize $^*$Note: Results are presented as "MAPE under anomalous data / MAPE under clean data (degradation scale)".}
\end{tabular}
\end{table*}

\begin{table*}[!ht]
\centering
\small 
\caption{Comparison of MAPE for moisture prediction.}
\label{Performance degradation (moisture)}
\setlength{\tabcolsep}{5pt} 
\begin{tabular}{llccccc}
\toprule
\textbf{Fabric} & \textbf{Model} & \multicolumn{5}{c}{\textbf{Percentage of Training Data}} \\
\cmidrule(lr){3-7}
\textbf{Type} & & \textbf{20\%} & \textbf{40\%} & \textbf{60\%} & \textbf{80\%} & \textbf{100\%} \\
\midrule
\textbf{Nylon} & Direct Learning & 9.96 / 8.21 & 9.00 / 7.64 & 8.70 / 7.13 & 7.86 / 6.99 & 7.66 / 6.90 \\
& & (+1.75) & (+1.36) & (+1.56) & (+0.89) & (+0.76) \\
\cmidrule{2-7}
& Transfer Learning & 6.17 / 5.94 & 6.53 / 5.97 & 6.70 / 6.00 & 6.85 / 6.13 & 7.03 / 6.13 \\
& & (+0.22) & (+0.56) & (+0.70) & (+0.72) & (+0.90) \\
\cmidrule{2-7}
& EDTL & \textbf{6.12 / 5.94} & \textbf{5.99 / 5.93} & \textbf{6.12 / 5.99} & \textbf{6.04 / 5.98} & \textbf{6.13 / 6.04} \\
& & \textbf{(+0.18)} & \textbf{(+0.06)} & \textbf{(+0.13)} & \textbf{(+0.06)} & \textbf{(+0.09)} \\
\midrule
\textbf{Polyester} & Direct Learning & 7.87 / 6.97 & 6.66 / 6.00 & 6.54 / 5.92 & 6.58 / 5.89 & 6.51 / 5.86 \\
& & (+0.90) & (+0.66) & (+0.62) & (+0.69) & (+0.65) \\
\cmidrule{2-7}
& Transfer Learning & 6.55 / 6.04 & 6.33 / 5.91 & 6.40 / 5.77 & 6.45 / 5.88 & 6.40 / 5.76 \\
& & (+0.51) & (+0.42) & (+0.63) & (+0.57) & (+0.64) \\
\cmidrule{2-7}
& EDTL & \textbf{5.34 / 5.29} & \textbf{5.42 / 5.33} & \textbf{5.52 / 5.39} & \textbf{5.65 / 5.46} & \textbf{5.78 / 5.50} \\
& & \textbf{(+0.05)} & \textbf{(+0.09)} & \textbf{(+0.13)} & \textbf{(+0.19)} & \textbf{(+0.28)} \\
\bottomrule
\addlinespace[1ex]
\multicolumn{7}{l}{\footnotesize $^*$Note: Results are presented as ``MAPE under anomalous data / MAPE under clean data (degradation scale)".}
\end{tabular}
\end{table*}

However, transfer learning performance usually relies on the quality of the pre-trained models obtained from the source domain. If the prediction errors of the pre-trained models are larger in the source domain (see the MAPEs of fabric width $D^{\rm s}$ for production line B in Table~\ref{table_pretrain}), the performance of the target model can be degraded, resulting in negative transfer. 
Figure~\ref{fig_comparison} shows the comparison of transferring pre-trained models from production lines A2 and B.
Transfer learning and EDTL inheriting the pre-trained model from production line A2 demonstrated excellent prediction performance because the pre-trained model of A2 had high prediction accuracy, and knowledge was effectively transferred to production line A1.
By contrast, transfer learning using a pre-trained model from production line B performed worse than a pre-trained model from production line A, leading to worse prediction accuracy. This result can be attributed to the low quality of the pre-training model from production line B and the discrepancy with production line A1. By using EDTL, the model inheriting the pre-trained model from B can achieve predictive performance comparable to that of the pre-trained model from A2. This observation highlights the robustness of the proposed EDTL, as the resulting predictive performance shows low sensitivity to the quality of the pre-trained models.

Additionally, we further validated the robustness of the proposed method. In the real industrial scenario, uncertainties such as aging equipment or environmental stress often lead to sensor monitoring anomalies. These anomalous data instances deviate from the expected patterns of regular data, making models trained on such anomalies prone to unexpected performance failures. In further experiments, we compared the performance degradation of the model trained on anomalous data. We randomly selected a portion of the target domain data and added noise to induce data fluctuations, simulating sensor data anomalies in real-world scenarios. For the data of the $jth$ feature added noise from Gaussian distribution can be expressed as:
\begin{equation}
        x_{ij}^{\prime}=x_{ij}+\epsilon_{ij}, \ \  \epsilon_{ij} \sim \\ \mathcal{N}(\mu_j, \sigma_j)
        \label{data add noise}
\end{equation}
To generate anomalous fluctuations in the data, the parameters of the Gaussian distribution should be adjusted based on the scale of each feature. In our setting, the mean $\mu_j$ was uniformly fixed at 0, the standard deviation $\sigma_j$ was set to 5\% of the average value of the feature, and 20\% of the target domain data was randomly selected and replaced with anomalous data.

Tables~\ref{Performance degradation (electricity)}, ~\ref{Performance degradation (moisture)}, ~\ref{Performance degradation (width)}, and~\ref{Performance degradation (weight)} respectively present the differences in predictive performance for the three models when trained on anomalous data versus clean data. Similarly, we conducted separate analyses for the four predicted targets ($E^t$, $M^t$, $W^t$, and $D^t$) and cases with different amounts of data usage. We observed that neural network models are susceptible to anomalous data, with their substantial predictive performance decreasing when trained on such data. However, the results show that the decline in performance is reduced in models trained by transfer learning, while training with the EDTL method further mitigates the performance degradation. This can be attributed to the pre-trained models that have learned stable feature representations from high-quality historical data. They provide prior knowledge that helps prevent the model from overfitting to anomalous data during fine-tuning in the target domain.  By combining multiple transferred models, EDTL not only smooths the variance of individual learners but also leverages the knowledge from different base learners. Since each learner may respond differently to anomalous data, ensemble learning integrates this diverse information to enhance the overall robustness of the model.

\subsection{Comparison with existing models}
We compared our proposed method with existing models. The comparison models include Random Forest, Adaboost, K-Nearest Neighbor (K-NN), and Stacking Regressor, which have been applied to modeling the drying process in the textile industry \cite{19taur}. Random Forest and Adaboost are tree-based algorithms that are extended into ensemble models using bagging and boosting algorithms, respectively. K-NN is an unsupervised learning method that relies on aggregating nearby neighbors based on sample distances to make predictions, without a parameter training process. Stacking Regressor is constructed using a stacking method with the three models mentioned above. Random Forest and K-NN are the base regressors, and Adaboost serves as the meta-regressor \cite{19taur}.

\begin{figure}[!ht]
\includegraphics[width=0.8\linewidth,clip,keepaspectratio]{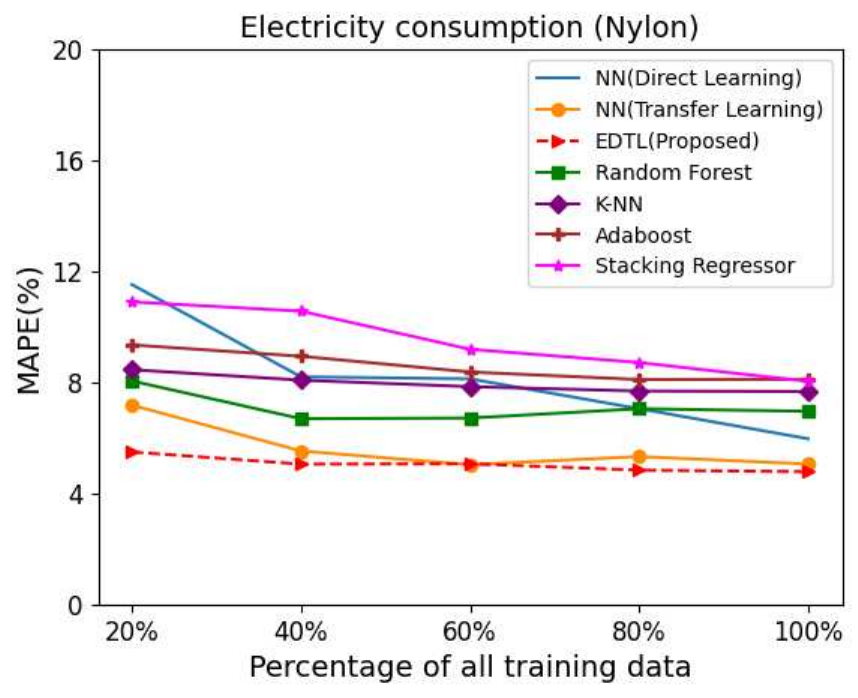}
\centering
\caption{Comparison of MAPE for different models on electricity consumption (Nylon, from A2 to A1).}\label{compare_ele_N}
\end{figure}

\begin{figure}[!ht]
\includegraphics[width=0.8\linewidth,clip,keepaspectratio]{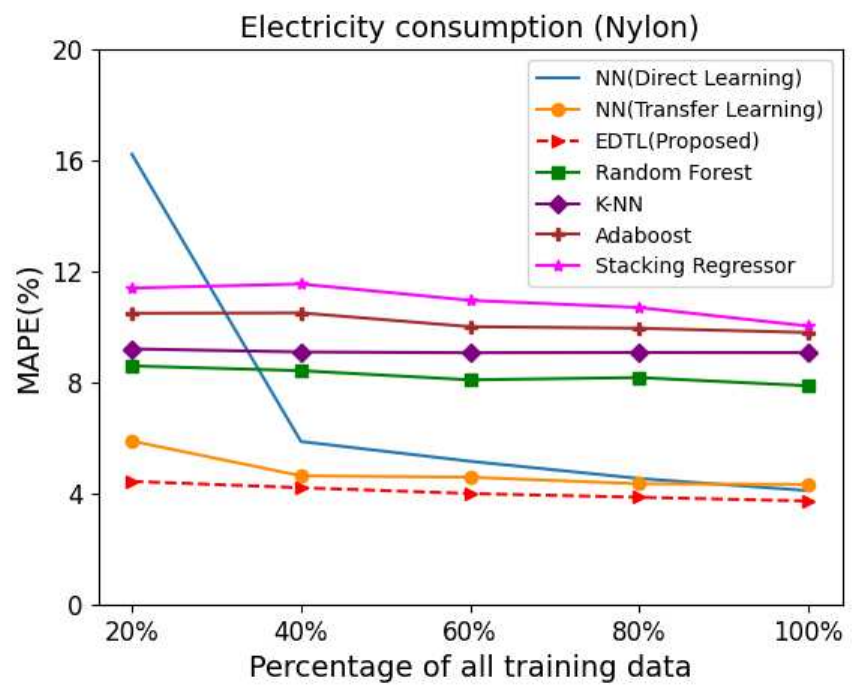}
\centering
\caption{Comparison of MAPE for different models on electricity consumption (Polyester, from A2 to A1).}\label{compare_ele_P}
\end{figure}

Since the SSM is the most energy-intensive equipment in the textile production process, identifying optimal energy parameters for it is a key priority. Therefore, we conducted a case study on electricity consumption prediction. Figures~\ref{compare_ele_N} and ~\ref{compare_ele_P} show the predictions of electricity consumption for the production of Nylon and Polyester, respectively. We observed that machine learning methods without neural networks are less sensitive to the amount of data used, with smaller variations in prediction error. However, their limited capacity to model nonlinear relationships makes it difficult to achieve acceptable performance. By contrast, methods based on neural networks demonstrate better capabilities in handling nonlinear relationships when sufficient data is available, achieving superior performance. Moreover, the use of transfer learning and ensemble learning further reduces their reliance on large datasets, broadening their applicability.

\begin{table*}[!ht]
\centering
\small 
\caption{Comparison of MAPE for fabric width prediction.}
\label{Performance degradation (width)}
\setlength{\tabcolsep}{5pt} 
\begin{tabular}{llccccc}
\toprule
\textbf{Fabric} & \textbf{Model} & \multicolumn{5}{c}{\textbf{Percentage of Training Data}} \\
\cmidrule(lr){3-7}
\textbf{Type} & & \textbf{20\%} & \textbf{40\%} & \textbf{60\%} & \textbf{80\%} & \textbf{100\%} \\
\midrule
\textbf{Nylon} & Direct Learning & 12.21 / 5.85 & 9.55 / 6.32 & 10.47 / 5.03 & 10.84 / 4.08 & 10.84 / 3.85 \\
& & (+6.36) & (+3.23) & (+5.44) & (+6.76) & (+6.99) \\
\cmidrule{2-7}
& Transfer Learning & \textbf{2.10 / 1.69} & 2.30 / 1.63 & 2.46 / 1.58 & 2.46 / 1.64 & 2.53 / 1.71 \\
& & \textbf{(+0.41)} & (+0.67) & (+0.88) & (+0.82) & (+0.82) \\
\cmidrule{2-7}
& EDTL & 1.92 / 1.28 & \textbf{1.93 / 1.30} & \textbf{2.04 / 1.34} & \textbf{2.06 / 1.37} & \textbf{2.04 / 1.37} \\
& & (+0.64) & \textbf{(+0.63)} & \textbf{(+0.70)} & \textbf{(+0.69)} & \textbf{(+0.68)} \\
\midrule
\textbf{Polyester} & Direct Learning & 11.02 / 7.44 & 6.83 / 3.23 & 6.17 / 1.47 & 5.60 / 1.06 & 4.83 / 0.93 \\
& & (+3.58) & (+3.60) & (+4.70) & (+4.54) & (+3.90) \\
\cmidrule{2-7}
& Transfer Learning & 1.36 / 0.75 & 1.18 / 0.67 & 1.08 / 0.61 & 1.16 / 0.63 & 1.14 / 0.62 \\
& & (+0.61) & (+0.51) & (+0.47) & (+0.53) & (+0.52) \\
\cmidrule{2-7}
& EDTL & \textbf{0.76 / 0.51} & \textbf{0.78 / 0.51} & \textbf{0.78 / 0.52} & \textbf{0.77 / 0.52} & \textbf{0.79 / 0.54} \\
& & \textbf{(+0.25)} & \textbf{(+0.27)} & \textbf{(+0.26)} & \textbf{(+0.25)} & \textbf{(+0.25)} \\
\bottomrule
\addlinespace[1ex]
\multicolumn{7}{l}{\footnotesize $^*$Note: Results are presented as ``MAPE under anomalous data / MAPE under clean data (degradation scale)".}
\end{tabular}
\end{table*}

\begin{table*}[!ht]
\centering
\footnotesize 
\caption{Comparison of MAPE for fabric weight per yard prediction.}
\label{Performance degradation (weight)}
\setlength{\tabcolsep}{5pt} 
\begin{tabular}{llccccc}
\toprule
\textbf{Fabric} & \textbf{Model} & \multicolumn{5}{c}{\textbf{Percentage of Training Data}} \\
\cmidrule(lr){3-7}
\textbf{Type} & & \textbf{20\%} & \textbf{40\%} & \textbf{60\%} & \textbf{80\%} & \textbf{100\%} \\
\midrule
\textbf{Nylon} & Direct Learning & 22.69 / 16.01 & 19.92 / 14.20 & 20.30 / 12.58 & 20.16 / 11.59 & 19.80 / 11.37 \\
& & (+6.68) & (+5.72) & (+7.72) & (+8.57) & (+8.43) \\
\cmidrule{2-7}
& Transfer Learning & 15.70 / 14.40 & 15.24 / 13.76 & 14.90 / 13.50 & 14.30 / 13.32 & 13.69 / 13.51 \\
& & (+1.30) & (+1.48) & (+1.40) & (+0.98) & (+0.18) \\
\cmidrule{2-7}
& EDTL & \textbf{12.60 / 12.48} & \textbf{11.93 / 11.72} & \textbf{10.47 / 10.21} & \textbf{10.32 / 10.04} & \textbf{10.41 / 10.07} \\
& & \textbf{(+0.12)} & \textbf{(+0.21)} & \textbf{(+0.26)} & \textbf{(+0.28)} & \textbf{(+0.34)} \\
\midrule
\textbf{Polyester} & Direct Learning & 20.12 / 14.32 & 21.47 / 13.49 & 21.25 / 13.11 & 20.46 / 12.84 & 19.50 / 12.49 \\
& & (+5.80) & (+7.98) & (+8.14) & (+7.62) & (+7.01) \\
\cmidrule{2-7}
& Transfer Learning & 19.74 / 13.62 & 18.59 / 11.92 & 16.50 / 10.98 & 14.40 / 10.93 & 13.27 / 10.72 \\
& & (+6.12) & (+6.67) & (+5.52) & (+3.47) & (+2.55) \\
\cmidrule{2-7}
& EDTL & \textbf{12.44 / 9.36} & \textbf{11.47 / 9.54} & \textbf{11.49 / 9.70} & \textbf{11.47 / 9.72} & \textbf{11.44 / 9.98} \\
& & \textbf{(+3.08)} & \textbf{(+1.93)} & \textbf{(+1.79)} & \textbf{(+1.75)} & \textbf{(+1.46)} \\
\bottomrule
\addlinespace[1ex]
\multicolumn{7}{l}{\footnotesize $^*$Note: Results are presented as ``MAPE under anomalous data / MAPE under clean data (degradation scale)".}
\end{tabular}
\end{table*}

\section{Conclusion}\label{sec_conclusion}

In this paper, we propose Ensemble Deep Transfer Learning (EDTL), a novel approach that integrates transfer learning with ensemble strategies and feature alignment to improve predictive accuracy and data efficiency in energy-intensive manufacturing. Our experiments, conducted on real-world textile factory data, demonstrate that EDTL improves prediction accuracy by 5.66\% and enhances model robustness by 3.96\% compared to conventional deep neural networks, particularly in data-limited scenarios (20\%–40\% data availability). These findings highlight EDTL’s capability to mitigate data scarcity challenges while maintaining reliable performance.

Beyond the textile industry, our method has broader practical implications for industrial decision-making, including applications in power grid management, chemical processing, and food production. For power grid companies, EDTL can enhance demand forecasting and energy optimization by enabling accurate predictions with minimal historical data. This is particularly valuable in dynamic environments where data availability is inconsistent. Similarly, industries that balance quality control and energy efficiency—such as chemical and food processing—can benefit from EDTL’s ability to adapt predictive models across different production settings, reducing operational costs while maintaining performance.

A key requirement for effective transfer learning is access to high-quality source domain data, yet sensor data fluctuations due to environmental and human factors introduce uncertainty in the pre-training process. While anomaly detection can mitigate low-quality data, it often results in excessive data loss. In future work, we aim to develop adaptive anomaly detection and uncertainty quantification techniques to enhance EDTL’s robustness and generalizability across diverse industrial applications. Further domain-specific investigations will also be conducted to refine the methodology for broader adoption in real-world manufacturing systems.

 \section*{Acknowledgments}
 During the preparation of this work the authors used ChatGPT for editing and grammar enhancement. After using this tool/service, the authors reviewed and edited the content as needed.

\clearpage

\bibliographystyle{IEEEtran}
\bibliography{IEEEabrv, mybib}

\end{document}